\pdfoutput=1
\documentclass[twocolumn]{article}

\usepackage{preprint}
\usepackage{amsmath,amssymb}
\usepackage{graphicx} 
\usepackage{siunitx}
\usepackage{caption}
\usepackage{subcaption}
\usepackage{tikz}
\usepackage{pgfplots}
\usepackage{xcolor}
\usepackage{grffile}
\usepackage[acronym]{glossaries}
\usepackage{booktabs}
\usepackage{multirow}
\usepackage{algorithm}
\usepackage{algpseudocode}
\usepackage{natbib}
\usepackage{authblk}
    
\pgfplotsset{compat=newest}
\usetikzlibrary{positioning,shapes,arrows,calc,backgrounds}
\usetikzlibrary{plotmarks}
\usepgfplotslibrary{patchplots}

\newacronym{acr:ltc}{LTC}{Low Temperature Combustion}
\newacronym{acr:rcci}{RCCI}{Reactivity Controlled Compression Ignition}
\newacronym{acr:ice}{ICE}{Internal Combustion Engine}
\newacronym{acr:com}{COM}{Control-oriented Model}
\newacronym{acr:pc}{PC}{Principal Component}
\newacronym{acr:pcd}{PCD}{Principal Component Decomposition}
\newacronym{acr:pca}{PCA}{Principal Component Analysis}
\newacronym{acr:gpr}{GPR}{Gaussian Process Regression}
\newacronym{acr:cpbc}{CPBC}{Cylinder Pressure-Based Control}
\newacronym{acr:ivc}{IVC}{Intake Valve Close}
\newacronym{acr:evo}{EVO}{Exhaust Valve Open}
\newacronym{acr:pfi}{PFI}{Port Fuel Injection}
\newacronym{acr:di}{DI}{Direct Injection}
\newacronym{acr:egr}{EGR}{Exhaust Gas Recirculation}
\newacronym{acr:se}{SE}{Square Exponential}
\newacronym{acr:rq}{RQ}{Rational Quadratic}
\newacronym{acr:ard}{ARD}{Automatic Relevance Determination}
\newacronym{acr:mae}{MAE}{Mean Absolute Error}
\newacronym{acr:icc}{ICC}{In-Cylinder Conditions}
\newacronym{acr:rms}{RMS}{Root Mean Square}
\newacronym{acr:doe}{DOE}{Design of Experiments}

\newacronym{acr:pso}{PSO}{Particle Swarm Optimization}
\newacronym{acr:bo}{BO}{Bayesian Optimisation}
\newacronym{acr:ei}{EI}{Expected Improvement}
\newacronym{acr:pi}{PI}{Probability of Improvement}

\newacronym{acr:gie}{GIE}{Gross Indicated Efficiency}
\newacronym{acr:itc}{ITC}{Idealised Thermodynamic Cycle}

\newcommand{\trans}{\mathsf{T}}
\DeclareMathOperator*{\argmax}{arg\,max}
\DeclareMathOperator*{\argmin}{arg\,min}

\date{\today}

\begin{document}
\title{Automated and Risk-Aware Engine Control Calibration Using Constrained Bayesian Optimization}

\author[1,\thanks{\texttt{m.g.vlaswinkel@tue.nl}}]{Maarten Vlaswinkel}    
\author[1]{Duarte Antunes}
\author[1,2]{Frank Willems}

\affil[1]{Control Systems Technology, Eindhoven University of Technology, Eindhoven, The Netherlands}
\affil[2]{Powertrains Department, TNO Mobility \& Built Environment, Helmond, The Netherlands}

\twocolumn[ %
  \begin{@twocolumnfalse}
    \maketitle

    \begin{abstract}
        Decarbonization of the transport sector sets increasingly strict demands to maximize thermal efficiency and minimize greenhouse gas emissions of Internal Combustion Engines. This has led to complex engines with a surge in the number of corresponding tuneable parameters in actuator set points and control settings. 
        Automated calibration is therefore essential to keep development time and costs at acceptable levels. In this work, an innovative self-learning calibration method is presented based on in-cylinder pressure curve shaping. This method combines Principal Component Decomposition with constrained Bayesian Optimization. To realize maximal thermal engine efficiency, the optimization problem aims at minimizing the difference between the actual in-cylinder pressure curve and an Idealized Thermodynamic Cycle. By continuously updating a Gaussian Process Regression model of the pressure's Principal Components weights using measurements of the actual operating conditions, the mean in-cylinder pressure curve as well as its uncertainty bounds are learned. This information drives the optimization of calibration parameters, which are automatically adapted while dealing with the risks and uncertainties associated with operational safety and combustion stability.
        This data-driven method does not require prior knowledge of the system.  
        The proposed method is successfully demonstrated in simulation using a Reactivity Controlled Compression Ignition engine model. The difference between the Gross Indicated Efficiency of the optimal solution found and the true optimum is \SI{0.017}{\%}. For this complex engine, the optimal solution was found after \SI{64.4}{s}, which is relatively fast compared to conventional calibration methods.
        
    \end{abstract}
	\vspace{0.35cm}
    \end{@twocolumnfalse}%
]

\section{Introduction}
The current demand on increasing the thermal efficiency and decreasing the emissions of \glspl{acr:ice} has led to the development of more advance combustion concepts \citep{Dempsey_2014}. \gls{acr:rcci} is one of the considered concepts \citep{Kokjohn_2011} for ultra efficient and clean combustion. \gls{acr:rcci} combustion is characterised by the use of two fuels. A low-reactivity fuel (e.g., E85) and a high-reactivity fuel (e.g., Diesel) are mixed in the combustion chamber. This makes it possible to control the reactivity of the fuel, thus giving control over the ignition timing.

The need to meet current and future increasingly strict emission legislations has led to an increase in the number of subsystems and actuators \citep{willems_2018,Norouzi_2021}. This leads to an increase in the number of tuneable actuator setpoints, references and feedback control parameters requiring tuning during \gls{acr:ice} control calibration. As the number of parameters increase, the time and complexity of the calibration process will rise to unacceptable levels. During calibration the settings for the parameters are determined that make sure that the engine meets pre-described performance targets. As a result, this will make \gls{acr:ice} calibration a more important part of \gls{acr:ice} development. Developing  novel calibration methods is a vital step to keep future and more advanced \glspl{acr:ice} commercially viable \citep{Paykani_2021}. In this work, the focus will be on the calibration of the fuel actuator settings related to timing and quantity of injection.

\subsection{Time Efficient Engine Calibration}
Calibration approaches have been proposed using model-based methods (e.g., \cite{Grasreiner_2017}) or using a predetermined \gls{acr:doe} (e.g., \cite{Motlagh_2020,xia_2020,Willems_2021,Biswas_2022,Moradi_2022}). Model-based methods use a validated combustion model to determine optimal parameter settings. These models are validate using experimental data. The quality of calibration is greatly determined by the quality of the model. \gls{acr:doe} methods treat the engine as a black-box system and use no, or only a small amount, of prior system knowledge. \gls{acr:doe} is used as a tool to determine key system behaviour in minimal testing time. It separate the collection of engine data and the process of determining the optimal actuator setpoints. The data collected using the \gls{acr:doe} is used as an input to a regression tool. 
\citet{xia_2020} was able to include cycle-to-cycle variations into the calibration problem. 

Developing an engine model or determining the \gls{acr:doe} are not trivial tasks. For experimental combustion concepts the safe operating regions are often unknown. Therefore, selecting appropriate actuator ranges is challenging and requires expertise. Simultaneous collecting data and the optimization of the \gls{acr:ice} performance can solve this challenge. \gls{acr:bo} is seen as a more adaptive and iterative method for \gls{acr:doe} \citep{Greenhill_2020}. The system is treated as a black-box, thus no prior knowledge of the system is required. For \glspl{acr:ice}, the safe operating domain is not linked to actuator ranges. Therefore, the \gls{acr:bo} approach should be aware of possible risky behaviour which has not been tackled in the literature, to the best of the authors knowledge.

\subsection{Shaping of the In-cylinder Pressure}
In this paper, an automated, risk-aware calibration method is presented. This method requires no prior knowledge of the system. The method is able to deal with unknown constraint boundaries and cycle-to-cycle variations. It extends upon the \gls{acr:pcd} and \gls{acr:gpr} based in-cylinder pressure model presented in \citet{vlaswinkel_2024} and the \gls{acr:itc} based cost function presented in \citet{vlaswinkel_2023}. The in-cylinder pressure model will predict the in-cylinder pressure between \gls{acr:ivc} and \gls{acr:evo} given actuator setpoints. The \gls{acr:itc} based cost function generates an idealised in-cylinder pressure profile. The calibration problem is reformulated into a probabilistic framework using a constrained \gls{acr:bo} approach.

\subsection{Outline of this Paper}
This work is organized as follows. In Section~\ref{sec:prob_desc}, a description of the system and calibration problem are presented. Section~\ref{sec:bof} presents the automated risk-aware calibration method. Sections~\ref{sec:pca}~until~\ref{sec:gpr} give a summary of the work presented in \citet{vlaswinkel_2023} and \citet{vlaswinkel_2024}. In Sections~\ref{sec:af}~until~\ref{sec:pso}, the new contributions to the calibration method are presented. Section~\ref{sec:results} discusses the observed behaviour of the method and a comparison is made between different acquisition functions using a simulated \gls{acr:rcci} engine fuelled with Diesel and E85.

\section{Problem Description} \label{sec:prob_desc}
The main objective of this work is to maximize thermal efficiency for a requested power output, while respecting bounds on mechanical stresses and combustion stability, by automated calibration of the engine control settings. 
 
\subsection{System Description}
  In this research, the optimization of the fuel settings of a single cylinder \gls{acr:rcci} engine is studied to demonstrate the potential of the novel automated control calibration method. This engine is equipped with a standard \gls{acr:di} injector, which injects Diesel directly into the cylinders.  The actuation signals of the \gls{acr:di} injector consist of the start-of-injection $\text{SOI}_\text{DI}$ and the injected mass of Diesel $m_\text{DI}$. For \gls{acr:rcci} research, an additional injector is installed in the intake for \gls{acr:pfi} of E85. The actuation signal of the port fuel injector consists of the mass of injected E85 $m_\text{PFI}$. The start-of-injection of the port fuel injector is kept constant at \num{-320} Crank Angle Degrees after Top Dead Center (CADaTDC). Note that we only consider single pulse \gls{acr:di} and \gls{acr:pfi} fuel injection. The cylinder intake and exhaust conditions associated with the air path $s_\text{air}$, such as \gls{acr:egr} ratio, intake and exhaust manifold pressure, and intake manifold temperature are kept constant in this work. The engine specifications and the operating conditions studied are listed in Table~\ref{tab:eng_prop}. The operating conditions are chosen to align with the model developed in \citet{vlaswinkel_2024}.

\subsection{Problem Formulation}
During the calibration of the combustion process, the fuelling parameters $s_\text{fuel}$ are used to maximise the gross indicated efficiency $\text{GIE}$ defined as
\begin{equation}
    \text{GIE}(s_\text{fuel},\,s_\text{air}) = \frac{\int_\Theta  p(\theta,\,s_\text{fuel},\,s_\text{air}) \,dV(\theta)}{Q_\text{fuel}},
\end{equation}
where $p(\theta)$ is the in-cylinder pressure over the crank-angle position $\theta \in \Theta$, $\Theta = [-180\,,180]\,\si{CADaTDC}$, $V(\theta)$ is the cylinder volume, and $Q_\text{fuel}$ is the total fuel energy. For the \gls{acr:rcci} engine studied, three fuelling parameters $s_\text{fuel}$ are defined. First, the total fuel energy
\begin{equation} \label{eq:Qfuel}
    Q_\text{fuel} = m_\text{PFI} \text{LHV}_\text{PFI} + m_\text{DI} \text{LHV}_\text{DI}
\end{equation}
with $m_\text{PFI}$ and $m_\text{DI}$ the injected fuel mass and $\text{LHV}_\text{PFI}$ and $\text{LHV}_\text{DI}$ the lower-heating value of the \acrfull{acr:pfi} and \acrfull{acr:di} fuels, respectively, is considered. Second, we use the energy-based blend ratio
\begin{equation}
    \text{BR} = \frac{m_\text{PFI} \text{LHV}_\text{PFI}}{Q_\text{fuel}}.
\end{equation}
And third, the start-of-injection of the directly injected fuel $\text{SOI}_\text{DI}$ is used.

Engine operation is typically defined by engine speed and power output. The gross Indicated Mean Effective Pressure $\text{IMEP}_\text{g}$ is a measure for the provided piston work during the power stroke and is given by
\begin{equation}
    \text{IMEP}_\text{g}(s_\text{fuel},\,s_\text{air}) = \frac{\int_\Theta  p(\theta,\,s_\text{fuel},\,s_\text{air}) \,dV(\theta)}{V_\text{d}}
\end{equation}
with displacement volume $V_\text{d}$. For a meaningful comparison of results, this value is kept constant during optimisation and should meet the driver's request $\text{IMEP}_\text{g,req}$. The engine speed is kept constant.
To ensure safety and stable combustion, the optimisation process must respect the limits of the maximum in-cylinder pressure $p_\text{ub}$, the increase rate of the maximum pressure $dp_\text{ub}$ and the coefficient of variation 
\begin{equation}
    \text{cov}(\text{IMEP}_\text{g}) = \frac{\sigma_{\text{IMEP}_\text{g}}}{\mu_{\text{IMEP}_\text{g}}}
\end{equation}
with $\mu_{\text{IMEP}_\text{g}}$ and $\sigma_{\text{IMEP}_\text{g}}$ the mean and standard deviation of $\text{IMEP}_\text{g}$, respectively.

In summary, the engine optimization problem can be formulated as:
\begin{subequations} \label{eq:gen_optProb} \begin{align}   
    s^*_\text{fuel} = \argmax_{s_\text{fuel} \in \mathcal{S}_\text{fuel}} \quad & \text{GIE}(s_\text{fuel},\,s_\text{air}), \label{eq:gen_optprob_gie}\\
    \text{s.t.} \quad & \text{IMEP}_\text{g}(s_\text{fuel},\,s_\text{air}) = \text{IMEP}_\text{g,req}, \label{eq:gen_optprob_imep}\\
    & \text{cov}\left(\text{IMEP}_\text{g}(s_\text{fuel},\,s_\text{air})\right) < \nonumber\\&\left[\text{cov}\left(\text{IMEP}_\text{g}(s_\text{fuel},\,s_\text{air})\right)\right]_\text{ub},\label{eq:gen_optprob_covimep}\\
    & \max_\theta(p(\theta,\,s_\text{fuel},\,s_\text{air})) < p_\text{ub}, \label{eq:gen_optprob_pmax}\\
    & \max_\theta\left(\frac{\partial p(s_\text{fuel},\,s_\text{air}))}{\partial \theta}\Bigg|_\theta\right) < dp_\text{ub}. \label{eq:gen_optprob_dpmax}
\end{align} \end{subequations}
where $s_{\text{fuel}}=[Q_\text{fuel},\,\text{BR},\,\text{SOI}_{\text{DI}}]^\text{T}$ and $s_{\text{fuel}}^*$ the optimal fuel parameters which solves (\ref{eq:gen_optProb}). The work generated by combustion is given by (\ref{eq:gen_optprob_imep}), the combustion stability limit by (\ref{eq:gen_optprob_covimep}) and the safety constraints by (\ref{eq:gen_optprob_pmax})-(\ref{eq:gen_optprob_dpmax}) to limit mechanical stresses. 

The system is treated like a black box. At the beginning of the optimization process, only an initial $s_\text{fuel}$ must be provided that meets the combustion stability and safety constraints of (\ref{eq:gen_optprob_covimep})-(\ref{eq:gen_optprob_dpmax}). 

\begin{table}
    \centering
    \caption{Properties of the simulated single cylinder \acrlong{acr:rcci} engine} \label{tab:eng_prop}
    \begin{tabular}{p{0.25\textwidth} p{0.15\textwidth}}
    \toprule
    Parameter & Value \\ \midrule
    PFI fuel & E85 \\
    DI fuel & Diesel \\
    Compression ratio & 17.2 \\
    Bore & \SI{130}{mm} \\
    Stroke & \SI{162}{mm} \\
    Displacement volume $V_\text{d}$ & \SI{2200}{mm^3} \\
    Engine speed & \SI{1200}{rpm} \\
    Intake manifold\newline pressure & \SI{1}{bar} \\
    Intake manifold\newline temperature & \SI{45}{\degree C} \\
    EGR ratio & \num{0.2} \\
    DI common-rail\newline pressure & \SI{600}{bar} \\
     \bottomrule
    \end{tabular}
\end{table}

\section{Bayesian Optimisation Framework} \label{sec:bof}
\begin{figure*}[!t]
    \centering
    \begin{subfigure}[b]{\textwidth}
        \centering
		\includegraphics{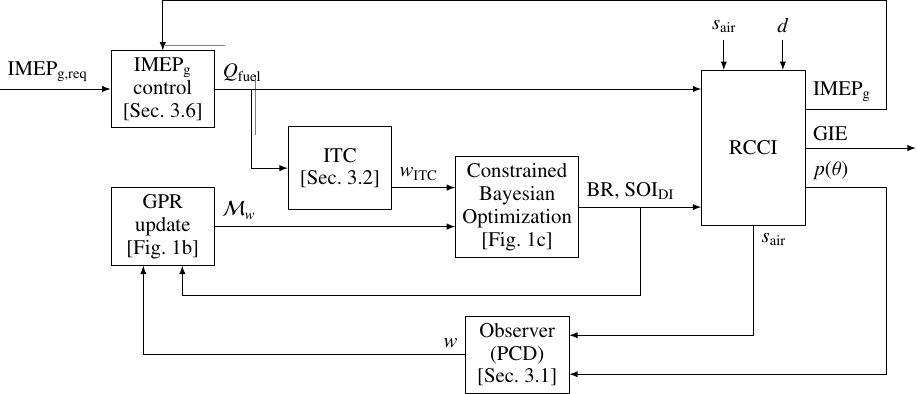}
        \caption{High-level overview of the Bayesian Optimization framework} \label{fig:general_schem}
    \end{subfigure}

    \begin{subfigure}[b]{\textwidth}
        \vspace{0.75em}
        \centering
		\includegraphics{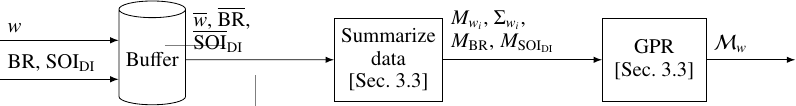}
        \caption{Updating of Gaussian Process Regression model; runs after $n$ combustion cycles after $\text{IMEP}_\text{g}$ has converged} \label{fig:gprUpdate_schem}
    \end{subfigure}
    
    \begin{subfigure}[b]{\textwidth}
        \vspace{0.75em}
        \centering
		\includegraphics{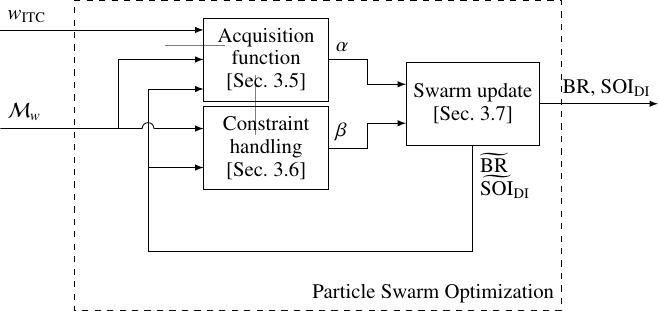}
        \caption{Bayesian optimization approach; runs after and update of the \acrfull{acr:gpr} update block until the \acrfull{acr:pso} has converged} \label{fig:bayesOpt_schem}
    \end{subfigure}
    \caption{Schematic representation of the optimisation approach presented in this work.} \label{fig:diag_schem}
\end{figure*}
In this section, the \acrfull{acr:bo} framework is presented that solves (\ref{eq:gen_optProb}) without the need for prior knowledge of the engine. This innovative framework is schematically represented in Figure~\ref{fig:diag_schem} and consists of four essential components:
\begin{enumerate}
    \item Next-cycle $\text{IMEP}_\text{g}$ control to realize the requested engine load by the driver;
    \item Targeted Ideal Thermodynamic Cycle (ITC);
    \item Learning a model of the weights $w$ associated with the Principal Components of the observed in-cylinder pressure;
    \item Constrained Bayesian Optimization to determine the fuel optimal $s^*_\text{fuel}$.
\end{enumerate}

Figure~\ref{fig:general_schem} gives an overview of the interaction between the different components in the framework. The \gls{acr:bo} is driven by a \acrfull{acr:pcd} of the observed in-cylinder pressure and a \acrfull{acr:gpr} model $\mathcal{M}_w$ that maps $\text{BR}$ and $\text{SOI}_\text{DI}$ to the \gls{acr:pc} weights $w$. To guarantee a power output equal to the driver's demand, the $\text{IMEP}_\text{g}$ is controlled outside the optimization loop by a next-cycle combustion controller using $Q_{\text{fuel}}$. To reduce the computation time of the \gls{acr:gpr} model, the collected data is summarized, as shown in Figure~\ref{fig:gprUpdate_schem}. After convergence of $\text{IMEP}_\text{g}$, the buffer will collect the \gls{acr:pc} weights of the last $n$ combustion cycles. The mean and standard deviation of these collected \gls{acr:pc} weights, i.e. $M_{wi}$ and $\Sigma_{wi}$ respectively, are added to the \gls{acr:gpr} model. This continuous update of the \gls{acr:gpr} model is also referred to as the self-learning part of the framework.
Figure~\ref{fig:bayesOpt_schem} shows the process of selecting a new actuator settings candidate. After new data has been added to the \gls{acr:gpr} model, a new solution is found for $\text{BR}$ and $\text{SOI}_\text{DI}$. An acquisition function is optimized using the current information in the \gls{acr:gpr} model while respecting the constraints. The optimization is carried out by a \gls{acr:pso}. After convergence of the \gls{acr:pso}, the optimal candidate that solves (\ref{eq:gen_optProb}) given the current information about the system is found and is applied to the system. The main components of the BO framework will be discussed in more detail in the following sections. 

\subsection{Principle Component Decomposition of In-Cylinder Pressure}  \label{sec:pca}
The current section is based upon the work described in \citet{vlaswinkel_2024}.
The sampled in-cylinder pressure $p(\theta,\,s_\text{fuel},\,p_\text{im})$ at crank angle $\theta \in \{-180,\,-180 + \Delta\text{CA},\, \dots, 180 - \Delta\text{CA},\, 180\}\,\si{CADaTDC}$ with $\Delta\text{CA}$ the crank angle resolution is decomposed as
\begin{equation} \label{eq:pcd_pressure}
    p(\theta,\,s_\text{fuel},\,p_\text{im}) = p_\text{mot}(\theta,\,p_\text{im}) + w(s_\text{fuel})^\trans f(\theta),
\end{equation}
where $w(s_\text{fuel}) \in \mathbb{R}^{n_\text{PC}}$ is a vector of weights and $f(\theta) \in \mathbb{R}^{n_\text{PC}}$ is the vector of \glspl{acr:pc} with $n_\text{PC}$ the number of \glspl{acr:pc}. The motored pressure is modelled by an adiabatic process and is given by
\begin{equation}
    p_\text{mot}(\theta,\,p_\text{im})  = p_\text{im} \left(\frac{V(\SI{-180}{CADaTDC})}{V(\theta)}\right)^\kappa
\end{equation}
with cylinder volume $V(\theta)$ and specific heat ratio $\kappa$. In the vectors $w(s_\text{fuel})$ and $f(\theta)$, the $i$th element is related to the $i$th \gls{acr:pc}. The in-cylinder condition $s_\text{fuel} \in \mathcal{S}$ are in the set $\mathcal{S}$ spanning the modelled operation domain. It is assumed that the in-cylinder pressure during the intake stroke is equal to the intake manifold pressure $p_\text{im}$.

The eigenvalue method is used to compute the \glspl{acr:pc}. The $n_\text{train}\cdot n_\text{cyc}$ in-cylinder pressures $p(\theta,\,s^*_\text{fuel})$ contained in the training set are used, where $n_\text{train}$ is the number of experiments in the training dataset and $n_\text{cyc}$ the number of combustion cycles within an experiment. The fuel setpoints for \gls{acr:pc} training are denoted by $s^*_\text{fuel} \in \mathcal{S}^* \subset \mathcal{S}$. The vector $F_i$ is the $i$th unit eigenvector of the matrix $P P^\trans$ where $P \in \mathbb{R}^{n_\text{CA} \times n_\text{train}n_\text{cyc}}$ with $n_\text{CA}$ the number of crank angle values. The elements in matrix $P$ are defined as
\begin{equation}
    [P]_{ab} := p( \theta_a,\, s^*_{\text{fuel},b} ) -  p_\text{mot}(\theta_a,\,s^*_{\text{fuel},b}),
\end{equation}
such that the $a$th row of $P$ contains the values of the in-cylinder pressure at the $a$th crank angle for all $s_\text{fuel}^* \in \mathcal{S}^*$ and the $b$th column of $P$ contains the full in-cylinder pressure at all $\theta \in \{-180,\,-180 + \Delta\text{CA},\, \dots,\, 180 - \Delta\text{CA},\, 180\}\,\si{CADaTDC}$ for the $b$th $s_\text{ICC}^*$.
The $i$th \gls{acr:pc} is defined as
\begin{equation} \label{eq:pcd_weight}
    f_i(\theta_a) = [F_i]_a.
\end{equation}

Using these \glspl{acr:pc}, the weight related to the $i$th \gls{acr:pc} is given by
\begin{equation}
    w_i(s_\text{fuel}) = P(s_\text{fuel}) F_i,
\end{equation}
where $[P(s_\text{fuel})]_a = p(\theta_a,\,s_\text{fuel}) - p_\text{mot}(\theta_a,\,p_\text{im})$.

The training set generates a single set of \glspl{acr:pc}. These \glspl{acr:pc} are ordered by the amount of variation in the dataset it captures, where $i=1$ is the \gls{acr:pc} that captures the largest variation. In \citet{vlaswinkel_2024}, it was determined that the first 8 \glspl{acr:pc} are required to properly describe the in-cylinder pressure using \gls{acr:pcd}.

\subsection{Idealised Thermodynamic Cycle Tracking} \label{sec:itc}
Using the \gls{acr:pcd} from the previous section, we introduce an efficiency-based performance measure, which determines the energy loss compared to a pre-defined ideal in-cylinder pressure curve. For more details, the reader is referred to \citet{vlaswinkel_2023}. 

\subsubsection{Energy Loss Criterion}
The total amount of energy has to be conserved during combustion. This means
\begin{equation} \label{eq:energybalance}
	W_\text{ind,g}  = \text{IMEP}_\text{g}  V_\text{d} = Q_\text{fuel} + Q_\text{loss},
\end{equation}
where $W_\text{ind,g} = \int p\, dV$ is the gross indicated work applied to the piston, $V_\text{d}$ is the displacement volume of the cylinder, $Q_\text{loss}$ are the total energy losses and $Q_\text{fuel}$ is the total fuel energy as defined in (\ref{eq:Qfuel}). These losses are divided into two parts: 1) the energy loss obtained when running an \acrfull{acr:itc} and 2) the additionally energy losses $Q_\text{NITC}$ associated with the actual non-\gls{acr:itc}. This can be written as
\begin{equation} \label{eq:energyloss}
	Q_\text{loss} = (1 - \eta_\text{ITC})Q_\text{fuel}  + Q_\text{NITC},
\end{equation}
where $\eta_\text{ITC}$ is the indicated fuel conversion efficiency of the \gls{acr:itc} process.
Substituting (\ref{eq:energyloss}) into (\ref{eq:energybalance}) gives
\begin{equation} \label{eq:Q_nitc}
	Q_\text{NITC} = W_\text{ind,g} - \eta_\text{ITC} Q_\text{fuel}.
\end{equation}
The \gls{acr:gie} is maximized when $Q_\text{NITC}$ is as close to zero as possible. This is done by defining the cost function
\begin{equation} \label{eq:energy_cost1}
    J = Q_\text{NITC}^2.
\end{equation}
Given the \gls{acr:pcd} of a measured in-cylinder pressure curve $p(\theta)$ and of a pressure curve  $p_\text{ITC}(\theta)$ associated with \gls{acr:itc}, (\ref{eq:energy_cost1}) can be rewritten using (\ref{eq:pcd_pressure}):
\begin{align}
	J &= 
		\left(\int_\Theta w(s_\text{fuel},\,s_\text{air})^\trans f(\theta) + p_\text{mot}(\theta,\,p_\text{im}) \ +\right. \nonumber \\ 
        &\quad\ \epsilon(\theta) \,dV - \int_\Theta w_\text{ITC}^\trans f(\theta) + p_\text{mot}(\theta,\,p_\text{im})\ + \nonumber \\
        &\quad\  \left. \epsilon_\text{ITC}(\theta) \,dV\right)^2, \nonumber \\
    &\begin{aligned}
   	\,=\ & \scalebox{0.75}{$( w(s_\text{fuel},\,s_\text{air}) - w_{\text{ITC}} )^\trans Z_1 (         w(s_\text{fuel},\,s_\text{air}) - w_{\text{ITC}} ) \ +$} \\
   	& \scalebox{0.9}{$2 ( w(s_\text{fuel},\,s_\text{air}) - w_{\text{ITC}} )^\trans Z_2 + Z_3$}
	\end{aligned} \label{eq:energy_cost2}
\end{align}
with
\begin{equation*}
	Z_1 = \iint_\Theta f(\theta_1) f^\trans(\theta_2) \,dV(\theta_1)\,dV(\theta_2),
\end{equation*}
\begin{equation*}
	Z_2 = \iint_\Theta f(\theta_1) ( \epsilon(\theta_2) - \epsilon_\text{ITC}(\theta_2) ) \,dV(\theta_1)\,dV(\theta_2)
\end{equation*}
and
\begin{equation*}
	\scalebox{0.75}{$Z_3 = \iint_\Theta \left( \epsilon(\theta_1) - \epsilon_\text{ITC}(\theta_1) \right)\left( \epsilon(\theta_2) - \epsilon_\text{ITC}(\theta_2) \right) \,dV(\theta_1)\,dV(\theta_2).$}
\end{equation*}
The weights $w$ and $w_\text{ITC}$ of the measured and \gls{acr:itc} curves are determined such that $\int_\Theta \epsilon(\theta)^2 \,dV(\theta)$ and $\int_\Theta \epsilon_\text{ITC}(\theta)^2 \,dV(\theta)$ are minimised, respectively. When enough \glspl{acr:pc} are used these terms will approach 0; therefore, it can be assumed that $Z_2 \approx 0$ and $Z_3 \approx 0$. This reduces (\ref{eq:energy_cost2}) to the following energy loss criterium:
\begin{equation} \label{eq:J_itc}
    \begin{aligned}
	J \approx J_\text{ITC} = & ( w(s_\text{fuel},\,s_\text{air}) - w_\text{ITC}(s_\text{fuel},\,s_\text{air}) )^\trans\ \cdot \\
    & Z_1 ( w(s_\text{fuel},\,s_\text{air}) - w_\text{ITC}(s_\text{fuel},\,s_\text{air}) ).
    \end{aligned}
\end{equation}

\subsubsection{Otto Cycle Pressure Model}   
In this article, the Otto cycle is used as \gls{acr:itc}, because it promises the highest thermal efficiency over a large operating range \citep{heywood_internal_2018}. It is a combination of two isentropic and two isochoric processes. It can be fully described by the total fuel energy $Q_\text{fuel}$, the pressure at the start of the compression stroke $p_\text{low}$, cylinder volume 
$V(\theta)$, and temperature and composition dependent specific heat ratio $\kappa$. The in-cylinder pressure corresponding to the Otto cycle is given by: 
\begin{equation} \label{eq:otto_cycle}
    p_\text{ITC}(\theta,\,Q_\text{fuel},p_\text{im}) =
	\scalebox{0.85}{$\begin{cases}
	   p_\text{low} \left(\frac{V(\SI{-180}{CAD})}{V(\theta)}\right)^\kappa,   & \text{if}\ \theta \in \Theta_\text{c}, \\
	   p_\text{high} \left(\frac{V(\SI{0}{CAD})}{V(\theta)}\right)^\kappa,   & \text{if}\ \theta \in \Theta_\text{e}
	\end{cases}$}
\end{equation}
with $\Theta_\text{c} = [-180,\,0]\,\si{CADaTDC}$, $\Theta_\text{e} = [0,\,180]\,\si{CADaTDC}$,
\begin{equation*}
	p_\text{high} = \frac{ \eta_\text{ITC} Q_\text{fuel}  - \int_{\SI{-180}{CAD}}^{\SI{0}{CAD}} p_\text{ITC}(\theta,\,Q_\text{fuel})\,dV}{ V^\kappa(\SI{0}{CAD}) \int_{\SI{-180}{CAD}}^{\SI{0}{CAD}} V^{-\kappa}(\theta) \,dV(\theta) }
\end{equation*}
and thermal efficiency of the Otto cycle $\eta_\text{ITC}$ is given by
\begin{equation} \label{eq:otto_eff}
	\eta_\text{ITC} = 1 - \frac{1}{r^{\kappa - 1}}
\end{equation}
with the compression ratio $r = \max(V) \min(V)^{-1}$. Using (\ref{eq:otto_eff}), it is found that the thermal efficiency increases when $\kappa$ increases. Note that the air-path conditions determine $p_\text{low}$ and $\kappa$.

\subsection{Gaussian Process Regression Prediction Model}  \label{sec:gpr}
\acrfull{acr:gpr} is used to predict the behavior of each component of $w(s_\text{fuel})$ over the entire operation domain $\mathcal{S}$. To include cycle-to-cycle variations in the in-cylinder pressure, $w(s_\text{fuel})$ is described by a stochastic process as
\begin{equation} \label{eq:gpr_w}
	w(s_\text{fuel}) := \mathcal{N}(\hat{w}(s_\text{fuel}),\,\hat{W}(s_\text{fuel}))
\end{equation}
with mean $\hat{w}(s_\text{fuel}) := \mathbb{E}[w(s_\text{fuel})]$ and variance $\hat{W}(s_\text{fuel}) := \mathbb{E}[ (w(s_\text{fuel}) - \hat{w}(s_\text{fuel})) \cdot (w(s_\text{fuel}) $\\$-\ \hat{w}(s_\text{fuel}))^\trans ]$. In this study, the correlation between output variable $w_i(s_\text{fuel})$ and $w_j(s_\text{fuel})$ $\forall i,\,j \in \{1,\,2,\,\dots,\,n_\text{PC}\}$ 
will be neglected (i.e., $\hat{W}(s_\text{fuel})$ is a diagonal matrix), since most literature on \gls{acr:gpr} assumes the output variables to be uncorrelated. This might affect the quality of the prediction of the cycle-to-cycle variation.

To improve the accuracy of prediction and the determination of hyperparameters, normalised in-cylinder conditions $\bar{s}_\text{fuel}$ and weights $\bar{w}_i(\bar{s}^*_\text{fuel})$ will be used. Scaling the in-cylinder condition uses the mean $\bar{\mu}_{s^*_\text{fuel},j}$ and standard deviation $\bar{\sigma}_{{s^*_\text{fuel}},j}$ of the $j$th in-cylinder condition variable over the full training set $\mathcal{S}^*$ as
\begin{equation}
    \bar{s}_{\text{fuel},j} = \frac{ s_{\text{fuel},j} - \bar{\mu}_{s^*_{\text{fuel}},j} }{ \bar{\sigma}_{s^*_{\text{fuel}},j} }.
\end{equation}
The scaling of the weights uses the mean $\bar{\mu}_{w_i}$ and standard deviation $\bar{\sigma}_{w^*_{i}}$ of the $i$th in-cylinder conditions variable over the full training set $S^*$ as
\begin{equation}
    \bar{w}_i(\bar{s}^*_\text{fuel}) = \frac{ w_i(\bar{s}^*_\text{fuel}) - \bar{\mu}_{w_i} }{ \bar{\sigma}_{w_i} }.
\end{equation}
Following \citet{rasmussen_2016}, the scaled expected value and scaled covariance matrix without correlation can be computed as:
 \begin{equation} \label{eq:mvgpr_exp}
    \begin{gathered}
    \hat{\bar{w}}_i(\bar{s}_\text{fuel}) = K( \bar{s}_\text{fuel},\,\bar{s}^*_\text{fuel},\,\phi )\ \cdot \\
    \left(K(\bar{s}^*_\text{fuel},\,\bar{s}^*_\text{fuel},\,\phi) + \bar{W}(s^*_\text{fuel})\right)^{-1} \bar{w}_i(\bar{s}^*_\text{fuel})
    \end{gathered}
\end{equation}
and
\begin{equation} \label{eq:mvgpr_covar}
    \scalebox{0.85}{$
    \begin{gathered}
    \hat{\bar{W}}_{ii}(\bar{s}_\text{fuel}) = K( \bar{s}_\text{fuel},\, \bar{s}_\text{fuel},\,\phi )\ - \\  K( \bar{s}_\text{fuel},\,\bar{s}^*_\text{fuel},\,\phi ) \left(K(\bar{s}^*_\text{fuel},\,\bar{s}^*_\text{fuel},\,\phi) + \bar{W}(s^*_\text{fuel})\right)^{-1}\ \cdot \\
    K^{\trans}( \bar{s}_\text{fuel} ,\, \bar{s}^*_\text{fuel},\,\phi),
    \end{gathered}$}
\end{equation}
where $K( \cdot,\,\cdot,\,\phi)$ is the Mat\'{e}rn kernel with $\nu = \tfrac{2}{3}$ and $\phi = \{\varphi_\text{f},\,\Phi_\text{l}\}$ are the kernel's hyperparameters. The elements in the kernel are given by
\begin{equation}
    \begin{gathered}
    k( x,\,y,\,\phi) = \varphi_\text{f}^2 \left( 1 + \sqrt{3} \rho(x,\,y) \right)\ \cdot \\
    \exp\left( -\sqrt{3} \rho(x,\,y) \right)
    \end{gathered}
\end{equation}
with
\begin{equation}
    \rho(x,\,y) = \sqrt{ (x - y)^\trans \Phi_\text{l}^{-2} (x - y) }.
\end{equation}
For more information on the selection of the kernel and hyperparameters, the interested reader is referred to \citet{vlaswinkel_2024}.

Finally, the scaled expected value and scaled covariance matrix are descaled to complete the description of (\ref{eq:gpr_w}). The descaled expected value is given by
\begin{equation}
    \hat{w}_i(\bar{s}_\text{fuel}) =
    \hat{\bar{w}}_i(\bar{s}_\text{fuel}) \bar{\sigma}_{w_i} +
    \bar{\mu}_{w_i}
\end{equation}
and the descaled covariance matrix is given by
\begin{equation}
    \hat{W}_{ii}(\bar{s}_\text{fuel}) = \hat{\bar{W}}_{ii}(\bar{s}_\text{fuel}) \bar{\sigma}_{w_i}.
\end{equation}

\subsection{Predicted Cost and Constraint Violation}
Given the \gls{acr:pcd} of the in-cylinder pressure in Section~\ref{sec:pca}, the introduced cost function $J_\text{ITC}$ associated with energy loss in Section~\ref{sec:itc},
and the \gls{acr:gpr} prediction model in Section~\ref{sec:gpr}, we can rewrite the optimization problem (\ref{eq:gen_optProb}) in terms of \glspl{acr:pc} $f(\theta)$ and weights $w(s_\text{fuel})$. Substituting (\ref{eq:pcd_pressure}) in (\ref{eq:gen_optProb}) results in the following formulation of the engine optimization problem: 
\begin{subequations} \label{eq:ref_optProb} \begin{align}   
    s^*_\text{fuel} = &\argmax_{s_\text{fuel} \in \mathcal{S}_\text{fuel}} \quad -J_\text{ITC}(s_\text{fuel},\,s_\text{air}), \label{eq:ref_optprob_gie}\\
    \text{s.t.} \quad & \hat{w}(s_\text{fuel})^\trans \int_\Theta f(\theta) \,dV(\theta) = \text{IMEP}_\text{g,req},  \label{eq:ref_optprob_imep}\\
    & \text{cov}\left(\text{IMEP}_\text{g}(s_\text{fuel},\,s_\text{air})\right) < \nonumber \\&  \left[\text{cov}\left(\text{IMEP}_\text{g}(s_\text{fuel},\,s_\text{air})\right)\right]_\text{ub}, \label{eq:ref_optprob_covimep}\\
    & \begin{gathered}p_\text{mot}(\theta,\,p_\text{im}) + \hat{w}(s_\text{fuel})^\trans f(\theta) < \\ p_\text{ub}\ \forall \theta \in \Theta,\end{gathered} \label{eq:ref_optprob_pmax}\\
    & \begin{gathered}\frac{\partial p_\text{mot}}{\partial\theta}\Bigg|_\theta + \hat{w}(s_\text{fuel})^\trans \frac{df}{d\theta}\Bigg|_\theta < \\ dp_\text{ub}\ \forall \theta \in \Theta. \end{gathered}\label{eq:ref_optprob_dpmax}
\end{align} \end{subequations}
Using the \gls{acr:gpr} model presented in Section~\ref{sec:gpr}, the predicted mean and variance of the parameters in (\ref{eq:ref_optProb}) can be determined. The required derivations will be presented in the following sections.

\subsection{Acquisition Function} \label{sec:af}
At the start of the optimization procedure, the cost function 
$J_\text{ITC}$($s_\text{fuel}$, $s_\text{air}$) in (\ref{eq:ref_optprob_gie}) is unknown. It is only possible to construct an estimate of the cost function given the previous observations up to time $k$ of collected in the set $\mathcal{D}_k$. The set $\mathcal{D}_k$ contains the applied $s_\text{fuel}$, corresponding observed $w(s_\text{fuel})$ and the best observed cost $J_\text{ITC,k}^*$ until time $k$. This description also includes uncertainty as a result from cycle-to-cycle behaviour and model uncertainty as a result of cycle-to-cycle variations and measurement noise. The acquisition function is a method of summarising the predicted cost and uncertainty into a scalar function. The acquisition function is used to drive the optimization.

The improvement of the cost is used as a measure to guide the optimization. The improvement at a point $s_\text{fuel}$ is defined as \citep[Ch. 7 and 8]{garnett_bayesoptbook_2023}:
\begin{equation} \label{eq:improvement}
    \begin{gathered}
    I(s_\text{fuel} \,|\, \mathcal{D}_k ) = \\
    \max\left( J_\text{ITC}( s_\text{fuel}\,|\, \mathcal{D}_k ) - J_\text{ITC,k}^*,\, 0 \right),
    \end{gathered}
\end{equation}
where $J_\text{ITC}( s_\text{fuel}\,|\, \mathcal{D}_k)$ is the mean predicted cost at $s_\text{fuel}$ given all information collected up to time $k$ denoted with set $\mathcal{D}_k$ and $J_{\text{ITC},k}^*$ is the best observed cost until time $k$. When the predicted cost is lower compared the best observed cost the improvement is equal to zero. Otherwise, the improvement is equal to the difference between the predicted and best observed cost. In this study, the two most common methods of using the improvement will be discussed: (1) \gls{acr:ei} and (2) \gls{acr:pi}. Both are formulated below for the case of noiseless as well as noisy observations.

\subsubsection{Expected Improvement}
The \gls{acr:ei} maximizes the expected value of the improvement defined in (\ref{eq:improvement}). The general formulation for the \gls{acr:ei} is given by: 
\begin{equation} \label{eg:af_gei}
    \begin{gathered}
    \alpha_\text{EI}( s_\text{fuel} \,|\, \mathcal{D}_k ) = \mathbb{E}\left[I(s_\text{fuel} \,|\, \mathcal{D}_k )\right].
    \end{gathered}
\end{equation}

In the case of noiseless observations this boils down to:
\begin{equation} \label{eq:alpha_ei}
    \begin{gathered}
     \alpha_\text{EI}( s_\text{fuel} \,|\, \mathcal{D}_k) = \int_0^{\infty} \left[\max\left(\tau - J^*_{\text{ITC},k},\,0 \right) \cdot\right. \\ \left.g_{\tilde{\mathcal{X}}^2}( \tau \,|\, s_\text{fuel},\,\mathcal{D}_k ) \right]\,d\tau,
    \end{gathered}
\end{equation}
where $g_{\tilde{\mathcal{X}}^2}: \mathbb{R}_{\geq0} \to (0,\,1)$ is the probability density function for a generalized chi-squared distribution. Using (\ref{eq:J_itc}), this distribution is parametrized as follows:
\begin{equation} \label{eq:chi2}
    \begin{gathered}
    J_\text{ITC}(s_\text{fuel}) = w(s_\text{fuel})^\trans Q_1 w(s_\text{fuel})\ + \\ q_2 w(s_\text{fuel}) + q_0
    \end{gathered}
\end{equation}
with $Q_2 = Z_1$, $q_1 = -2 w_\text{ITC}^\trans Z_1 w(s_\text{fuel})$ and $q_0 = w_\text{ITC}^\trans Z_1 w_\text{ITC}$. The mean and variance of the normal distributed $w(s_\text{fuel})$ are computed using the model described in Section~\ref{sec:gpr}.

When the observations are noisy, (\ref{eg:af_gei}) becomes:
\begin{equation} \label{eq:af_eino}
    \scalebox{0.9}{$\begin{gathered}
    \alpha_\text{EI-NO}( s_\text{fuel} \,|\, \mathcal{D}_k ) =
    \int_0^{\infty} \Big[ g_{\tilde{\mathcal{X}}^2}( \tau^* \,|\, s^*_{\text{fuel},k},\,\mathcal{D}_k ) \ \cdot \\
    \left. \int_0^{\infty} \left[\max\left(\tau - \tau^*,\,0 \right) g_{\tilde{\mathcal{X}}^2}( \tau \,|\, s_\text{fuel},\,\mathcal{D}_k ) \right]\,d\tau\right]\,d\tau^*,
    \end{gathered}$}
\end{equation}
here $\tau^*$ replaces $J_\text{ITC,k}^*$ in the original definition of $\alpha_\text{EI}$ in (\ref{eq:alpha_ei}) and $s^*_{\text{fuel},k}$ are the fuel conditions that realized the best observed cost until time $k$. The probability distribution related to the cost at $s^*_{\text{fuel},k}$ is computed using (\ref{eq:chi2}).

\subsubsection{Probability of Improvement}
The \gls{acr:pi} maximizes the probability of improving the system. The general formulation for the \gls{acr:pi} is given by: 
\begin{equation} \label{eq:af_gpi}
    \begin{gathered}
    \alpha_\text{PI}( s_\text{fuel} \,|\, \mathcal{D}_k ) = \\ 
    \text{Pr}\left( J_\text{ITC}( s_\text{fuel}\,|\, \mathcal{D}_k ) > J_\text{ITC,k}^* \,|\, \mathcal{D}_k \right).
    \end{gathered}
\end{equation}
In the case of noiseless observations this boils down to:
\begin{equation} \label{eq:af_pi}
    \begin{gathered}
     \alpha_\text{PI}( s_\text{fuel} \,|\, \mathcal{D}_k,\, J_\text{ITC,k}^* ) = \\
     1 - G_{\tilde{\mathcal{X}}^2}\left( J^*_{\text{ITC},k} \,|\, s_\text{fuel},\,\mathcal{D}_k \right),
     \end{gathered}
\end{equation}
where $G_{\tilde{\mathcal{X}}^2}: \mathbb{R}_{\geq0} \to (0,\,1)$ is the cumulative distribution function for a generalized chi-squared distribution parametrized according to (\ref{eq:chi2}) and the model presented in Section~\ref{sec:gpr}.

When the observations are noisy, (\ref{eq:af_gpi}) becomes:
\begin{equation} \label{eq:af_pino}
    \begin{gathered}
    \alpha_\text{PI-NO}( s_\text{fuel} \,|\, \mathcal{D}_k ) =
    \int_{0}^{\infty}  g_{\tilde{\mathcal{X}}^2}\left( \tau^* \,|\, s^*_\text{fuel},\, \mathcal{D}_k \right) \cdot \\
    \left( 1 - G_{\tilde{\mathcal{X}}^2}\left( \tau^* \,|\, s_\text{fuel},\,\mathcal{D}_k \right) \right) \,d\tau^*,
    \end{gathered}
\end{equation}
where $\tau^*$ replaces $J_\text{ITC,k}^*$ in the original definition of $\alpha_\text{PI}$ in (\ref{eq:af_pi}).

\subsection{Constraint Handling} \label{sec:constraint}
Similar to \citet{gelbart_2014}, a probabilistic view of the constraints will be used. The predicted probability of violating the constraints can be computed as non-independent events of violating each individual constraint. Given an input $s_\text{fuel}$, the predicted probability of violating each individual constraint is
\begin{equation}
    \scalebox{0.9}{$\tilde{\beta}_i(s_{\text{fuel},k}) = \text{Pr}\left( h_i(s_{\text{fuel},k} ) > 0  \,|\, s_{\text{fuel},k}\ \text{and}\ \mathcal{D}_k \right),$}
\end{equation}
where $h_i(s_{\text{fuel},k})$ with $i \in \{1,\,2,\,\dots,\,4\}$ are a reformulation of the constraints in (\ref{eq:ref_optprob_imep}) to (\ref{eq:ref_optprob_dpmax}). The reformulation for each constrained is discussed below.
The probability $\beta_{n_\text{const}}$ of violating the constraints, where $n_\text{const}$ is the number of constraints, can be recursively solved using
\begin{equation} \label{eq:beta}
    \begin{gathered}
    \beta_i(s_{\text{fuel},k}) = \beta_{i-1}(s_{\text{fuel},k}) (1 - \tilde{\beta}_i(s_{\text{fuel},k}))\ + \\
    \tilde{\beta}_i(s_{\text{fuel},k})
    \end{gathered}
\end{equation}
with $i \in \{0,\,1,\,\dots,\,n_\text{const}\}$ and $\beta_0(s_{\text{fuel},k}) = 0$. A $s_{\text{fuel},k}$ will be labelled as infeasible if the probability of violating the constraints exceeds 5\%. This limit can be increased to allow for more risky behaviour or decreased to be more risk averse.

It is assumed that the probability distribution of each constraint can be represented with a Gaussian probability distribution. To compute $\tilde{\beta}_i(s_{\text{fuel},k})$ the mean and variance of each constraint is required. This is computed using the current system knowledge $\mathcal{D}_k$. 

\subsubsection{$\text{IMEP}_\text{g}$ Constraint}
The constraint related to $\text{IMEP}_\text{g}$ and $\text{cov}(\text{IMEP}_\text{g})$ are given in (\ref{eq:ref_optprob_imep}) and (\ref{eq:ref_optprob_covimep}), respectively. Given the feedback controller regulating $Q_\text{fuel}$ to reach $\text{IMEP}_\text{g,req}$ satisfies (\ref{eq:ref_optprob_imep}) if $\text{IMEP}_\text{g,req}$ is reachable at the current $\text{BR}$ and $\text{SOI}_\text{DI}$. The reachability of $\text{IMEP}_\text{g,req}$ and the constraint of (\ref{eq:ref_optprob_covimep}) are combined as a constraint for the lower and upper bound on $\text{IMEP}_\text{g}$. The lower bound is given by
\begin{equation}
    \begin{gathered}
    h_1(s_{\text{fuel},k}) = w(s_{\text{fuel},k}) \int_{\SI{-180}{\degree}}^{\SI{180}{\degree}} f(\theta)\, dV(\theta)\ - \\
    \text{IMEP}_\text{g,req} \left(1 - \tfrac{1}{2}[\text{cov}(\text{IMEP}_\text{g})]_\text{ub} \right)
    \end{gathered}
\end{equation}
with $[\text{cov}(\text{IMEP}_\text{g})]_\text{ub}$ to upper bound of the coefficient of variation on $\text{IMEP}_\text{g}$. In this work, $[\text{cov}(\text{IMEP}_\text{g})]_\text{ub}$ is set to 5\%. Using the \gls{acr:gpr} model presented in Section~\ref{sec:gpr}, the mean and variance for the lower bound can be predicted using, respectively:
\begin{equation}
    \begin{gathered}
    \mu_{h_1}(s_{\text{fuel},k}\,|\,\mathcal{D}_k) = \\ 
    \hat{w}(s_{\text{fuel},k}\,|\,\mathcal{D}_k)^\trans \int_{\SI{-180}{\degree}}^{\SI{180}{\degree}} f(\theta)\, dV(\theta)\ - \\ 
    \text{IMEP}_\text{g,req} \left( 1 - \tfrac{1}{2}[\text{cov}(\text{IMEP}_\text{g})]_\text{ub} \right)
    \end{gathered}
\end{equation}
and
\begin{equation}
    \begin{gathered}
    \sigma^2_{h_1}(s_{\text{fuel},k}\,|\,\mathcal{D}_k) = \\
    \left(\int_{\SI{-180}{\degree}}^{\SI{180}{\degree}} f(\theta)\, dV(\theta)\right)^\trans \hat{W}(s_{\text{fuel},k}\,|\,\mathcal{D}_k)\ \cdot \\
    \left(\int_{\SI{-180}{\degree}}^{\SI{180}{\degree}} f(\theta)\, dV(\theta)\right).
    \end{gathered}
\end{equation}

The upper bound is given by
\begin{equation}
    \begin{gathered}
    h_2(s_{\text{fuel},k}) = w(s_{\text{fuel},k}) \int_{\SI{-180}{\degree}}^{\SI{180}{\degree}} f(\theta)\, dV(\theta)\ - \\
    \text{IMEP}_\text{g,req} \left(1 + \tfrac{1}{2}[\text{cov}(\text{IMEP}_\text{g})]_\text{ub} \right).
    \end{gathered}
\end{equation}
The mean and variance for the upper bound are computed as
\begin{equation}
    \begin{gathered}
    \mu_{h_2}(s_{\text{fuel},k}\,|\,\mathcal{D}_k) =  \\
    -\hat{w}(s_{\text{fuel},k}\,|\,\mathcal{D}_k)^\trans \int_{\SI{-180}{\degree}}^{\SI{180}{\degree}} f(\theta)\, dV(\theta)\ +\\
    \text{IMEP}_\text{g,req} \left( 1 + \tfrac{1}{2}[\text{cov}(\text{IMEP}_\text{g})]_\text{ub} \right)
    \end{gathered}
\end{equation}
and
\begin{equation}
    \sigma^2_{h_2}(s_{\text{fuel},k}\,|\,\mathcal{D}_k) = \sigma^2_{h_1}(s_{\text{fuel},k}\,|\,\mathcal{D}_k),
\end{equation}
respectively.

\subsubsection{Safety Constraints}
The safety constraint dealing with $\max_{\theta \in \Theta}(p(\theta,\,s_\text{fuel}))$ and $\max_{\theta \in \Theta}\left(\left.\frac{\partial p}{\partial \theta}\right|_{\theta,\,s_\text{fuel}}\right)$ are given in (\ref{eq:ref_optprob_pmax}) and (\ref{eq:ref_optprob_dpmax}), respectively. The constraint in (\ref{eq:ref_optprob_pmax}) is evaluated at $\theta_{p_\text{max}} = \argmax_{\theta \in \Theta} (\hat{w}(s_{\text{fuel},k}\,|\,\mathcal{D}_k)^\trans f(\theta) + p_\text{mp}(\theta))$ such that
\begin{equation}
    \begin{gathered}
    h_3(s_{\text{fuel},k}) = w(s_{\text{fuel},k}) f(\theta_{p_\text{max}})\ +\\ p_\text{mp}(\theta_{p_\text{max}}) -  p_\text{ub}.
    \end{gathered}
\end{equation}
Using the \gls{acr:gpr} model presented in Section~\ref{sec:gpr}, the mean and variance are computed as
\begin{equation}
    \begin{gathered}
    \mu_{h_3}(s_{\text{fuel},k}\,|\,\mathcal{D}_k) = \\
    \hat{w}(s_{\text{fuel},k}\,|\,\mathcal{D}_k)^\trans f(\theta_{p_\text{max}}) + p_\text{mp}(\theta_{p_\text{max}}) - p_\text{ub}
    \end{gathered}
\end{equation}
and
\begin{equation}
    \begin{gathered}
    \sigma^2_{h_3}(s_{\text{fuel},k}\,|\,\mathcal{D}_k) = \\
    f^\trans(\theta_{p_\text{max}}) \hat{W}_{\hat{w}}(s_{\text{fuel},k}\,|\,\mathcal{D}_k)f(\theta_{p_\text{max}}).
    \end{gathered}
\end{equation}
The constraint given in (\ref{eq:ref_optprob_dpmax}) is evaluated at $\theta_{dp_\text{max}} = \argmax_{\theta \in \Theta} (\hat{w}(s_{\text{fuel},k}\,|\,\mathcal{D}_k)^\trans \tfrac{df}{d\theta} + \frac{dp_\text{mp}}{d\theta})$ such that
\begin{equation}
    \begin{gathered}
    h_4(s_{\text{fuel},k}) = w(s_{\text{fuel},k}) \left.\tfrac{df}{d\theta}\right|_{\theta_{dp_\text{max}}}\ +\\ \left.\frac{dp_\text{mp}}{d\theta}\right|_{\theta_{dp_\text{max}}} - dp_\text{ub}
    \end{gathered}
\end{equation}
The mean and variance are computed as
\begin{equation}
    \begin{gathered}
    \mu_{h_4}(s_{\text{fuel},k}\,|\,\mathcal{D}_k) = \\
    \hat{w}(s_{\text{fuel},k}\,|\,\mathcal{D}_k)^\trans \left.\tfrac{df}{d\theta}\right|_{\theta_{dp_\text{max}}}\ +\\
    \left.\frac{dp_\text{mp}}{d\theta}\right|_{\theta_{dp_\text{max}}} - dp_\text{ub}
    \end{gathered}
\end{equation}
and
\begin{equation}
    \begin{gathered}
    \sigma^2_{h_4}(s_{\text{fuel},k}\,|\,\mathcal{D}_k) = \\
    \left(\left.\frac{df}{d\theta}\right|_{\theta_{dp_\text{max}}}\right)^\trans \hat{W}_{\hat{w}}(s_{\text{fuel},k}\,|\,\mathcal{D}_k) \ \cdot \\ 
    \left(\left.\frac{df}{d\theta}\right|_{\theta_{dp_\text{max}}}\right).
    \end{gathered}
\end{equation}

\subsection{Particle Swarm Optimisation with Constraints} \label{sec:pso}
The \gls{acr:bo} problem is non-convex and has multiple local optima. \gls{acr:pso} is used to find the global optimum of the acquisition function given the system knowledge $\mathcal{D}_k$. It is performed every iteration of the \gls{acr:bo} approach and is responsible for selecting the next solution being applied to the system. It uses different candidate solutions that over different iteration $\ell$ are moving towards the global optimum. Each candidate solution has a cost value, constraint violation probability, position and velocity. The computation of the cost value is discussed in Section~\ref{sec:af} and the constraint violation probability in Section~\ref{sec:constraint}. In this section, the updating rules of the position $s_\text{fuel}$ of the candidate solution is discussed. 

Several iterations of the \gls{acr:pso} are required to find the global optimum. A single iteration is denoted with $\ell \in \mathbb{N}_0$. The \gls{acr:pso} uses $n_\text{PSO}$ particles. Each particle has a location $\tilde{s}^i_{\text{fuel}} \in \mathcal{S}_\text{fuel}$ and velocity $v^i \in \mathbb{R}^2$, where the subscript $i \in \{1,\,2,\,\dots,\,n_\text{PSO}\}$ indicates the specific particle. Each iteration the location of all the particles is update according to
\begin{equation} \label{eq:siell}
    \tilde{s}^i_{\text{fuel},\ell} = \tilde{s}^i_{\text{fuel},\ell-1} + v^i_\ell
\end{equation}
with
\begin{equation} \label{eq:viell}
    \begin{gathered}
    v^i_\ell = c_0 v^i_{\ell - 1}\ + \\
    c_1 X_{1,\ell} ( \tilde{s}^i_{\text{fuel,p.best} } - \tilde{s}^i_{\text{fuel},\ell-1} )\ + \\
    c_2 X_{2,\ell} ( \tilde{s}_{\text{fuel,g.best}} - \tilde{s}^i_{\text{fuel},\ell-1} ),
    \end{gathered}
\end{equation}
where $\tilde{s}^i_{\text{fuel,p.best}}$ is the best observed cost of the $i$th particle, $\tilde{s}_{\text{fuel,g.best}}$ is the best observed cost over all particles, $c_k \in (0,\,1),\ k \in \{1,\,2,\,3\}$ are tuning parameters and $X_{j,\ell},\ j \in \{1,\,2\}$ are samples taken from the standard uniform distribution. The initial particle locations $\tilde{s}^i_{\text{fuel},0}$ are selected from a uniform distribution spanning the actuator range and initial particle velocities $v^i_0$ are selected from a normal distribution. 

A similar approach to select the location of the best cost as \citet{lampinen_2002} will be used. In their work, the following method selects $\tilde{s}^i_{\text{fuel,p.best}}$ and $\tilde{s}_{\text{fuel,g.best}}$ as a the best candidate solutions from the whole population when:
\begin{itemize}
    \item if $\tilde{s}_{\text{fuel},\ell}$ is feasible and the value of the objective function gives the largest or equal value of the previous observed costs with a feasible solution, or
    \item if $\tilde{s}_{\text{fuel},\ell}$ is feasible while the other previously observed solutions of the population are infeasible, or
    \item all solutions in the population are infeasible, but $\tilde{s}_{\text{fuel},\ell}$ has the lowest or equal violation of the constraints.
\end{itemize}
The constrained \gls{acr:pso} is summarized in Algorithm~\ref{alg:pso}.

\begin{algorithm*}[!t]
\caption{Overview of the constrained Particle Swarm Optimizer presented in Section~\ref{sec:pso}} \label{alg:pso}
\begin{algorithmic}[1]
    \State $\ell \gets 0$
    \State $\tilde{s}_{\text{fuel},0} \gets$ Uniform grid of $n_\text{PSO}$ particles spanning $\mathcal{S}_\text{fuel}$
    \State $v_0 \gets$ $n_\text{PSO}$ randomly generated velocities vectors
    
    \For{$\ell < 100$}
        \State $\ell \gets \ell + 1$  
        \ForAll{$\tilde{s}^i_{\text{fuel},\ell-1} \in \tilde{s}_{\text{fuel},\ell-1}$}
            \State $v^i_\ell \gets$ (\ref{eq:viell})
            \State $\tilde{s}^i_{\text{fuel},\ell} \gets$ (\ref{eq:siell})

            \State $\alpha(\tilde{s}^i_{\text{fuel},\ell} \,|\, \mathcal{D}_k) \gets$ (\ref{eq:alpha_ei})/(\ref{eq:af_eino})/(\ref{eq:af_pi})/(\ref{eq:af_pino})
            \State $\beta_{n_\text{const}}(\tilde{s}^i_{\text{fuel},\ell} \,|\, \mathcal{D}_k) \gets$ (\ref{eq:beta})

            \If{$\beta_{n_\text{const}}(\tilde{s}^i_{\text{fuel},\ell}) \leq 0.05$ and $\beta_{n_\text{const}}(\tilde{s}^i_{\text{fuel},\text{p.best}}) \leq 0.05$} \Comment{Feasibility checks}
                \State $\tilde{s}^i_{\text{fuel},\text{p.best}} \gets \argmax\left(\alpha(\tilde{s}^i_{\text{fuel},\text{p.best}}),\,\alpha(\tilde{s}^i_{\text{fuel},\ell})\right)$
            \ElsIf{$\beta_{n_\text{const}}(\tilde{s}^i_{\text{fuel},\ell}) \leq 0.05$ and $\beta_{n_\text{const}}(\tilde{s}^i_{\text{fuel},\text{p.best}}) > 0.05$}
                \State $\tilde{s}^i_{\text{fuel},\text{p.best}} \gets \tilde{s}^i_{\text{fuel},\ell}$
            \Else
                 \State $\tilde{s}^i_{\text{fuel},\text{p.best}} \gets \argmin\left(\beta_{n_\text{const}}(\tilde{s}^i_{\text{fuel},\text{p.best}}),\,\beta_{n_\text{const}}(\tilde{s}^i_{\text{fuel},\ell})\right)$
            \EndIf
        \EndFor
        \State $\tilde{s}_{\text{fuel},\text{g.best}} \gets \argmax\left(\alpha(\tilde{s}_{\text{fuel},\text{g.best}}),\,\alpha(\tilde{s}_{\text{fuel},\text{p.best}})\right)$ \Comment{Only use $\tilde{s}^i_{\text{fuel},\ell}$ with $\beta_{n_\text{const}}(\tilde{s}^i_{\text{fuel},\ell}) \leq 0.05$}
    \EndFor
    \State Apply $\tilde{s}_{\text{fuel},\text{g.best}}$ to the system
\end{algorithmic}
\end{algorithm*}

\section{Results} \label{sec:results}
\begin{table}
    \centering
    \caption{Settings for the simulated engine calibration in Section~\ref{sec:results} using the method presented in this study.} \label{tab:sim_settings}
    \begin{tabular}{p{0.15\textwidth} p{0.25\textwidth}}
        \toprule
        Parameter & Value\\ \midrule
        $\text{IMEP}_\text{g,ref}$ & \SI{4}{bar} \\
        $Q_\text{fuel}$ range & $[1639.6,\,2405.8]$\,\si{J/cycle} \\     
        BR range & $[0.7046,\,0.8188]$ \\
        $\text{BR}_0$ & \num{0.8} \\
        $\text{SOI}_\text{DI}$ range & $[-75,\,-35]$\,\si{CADaTDC} \\
        $\text{SOI}_{\text{DI},0}$ & \SI{-45}{CADaTDC} \\
        $[\text{cov}(\text{IMEP}_\text{g})]_\text{ub}$ & 10\%\\
        $p_\text{max}$ & \SI{200}{bar} \\
        $dp_\text{max}$ & \SI{25}{bar \per CAD} \\
        $n_\text{sample}$ & \num{25} \\
        $n_\text{PSO}$ & \num{100} \\
        $c_0$ & \num{0.1} \\
        $c_1$ & \num{0.01} \\
        $c_2$ & \num{0.1} \\
        \bottomrule
    \end{tabular}
\end{table}
In this section, the proposed method is evaluated from simulation results. For the engine described in Section \ref{sec:prob_desc}, a stochastic \gls{acr:rcci} engine model was developed in earlier work. Based on fuelling settings $s_\text{fuel}$ and air path conditions $s_\text{air}$, this engine model provides the in-cylinder pressure trace, including the cycle-to-cycle variation during the compression and expansion stroke. This model was validated using experimental engine data and shows good prediction capabilities. It is further described in \citet{vlaswinkel_2024}. 

Using this model, first a brief discussion of the exploration mechanics using $\alpha_\text{EI}$ is presented. Second, a comparison is made between the four acquisition functions presented in Section~\ref{sec:af}. The initial conditions and settings for the evaluation are given in Table~\ref{tab:sim_settings}. The actuator ranges and $\text{IMEP}_\text{g,ref}$ are chosen to be inline with the simulation model. The bounds on the inequality constraints ($p_\text{max}$, $dp_\text{max}$ an $[\text{cov}(\text{IMEP}_\text{g})]_\text{ub}$) are determined by the engine specifications. The parameters $n_\text{sample}$ and $n_\text{PSO}$ are determined using a trail-and-error approach. The \gls{acr:pso} parameters $c_0$, $c_1$ and $c_2$ are chosen to prefer a move towards the global optimum. 

\subsection{Evaluation of Single Acquisition Function}
\begin{figure*}
    \centering
    \begin{subfigure}[b]{0.45\textwidth}
        \centering
        \includegraphics{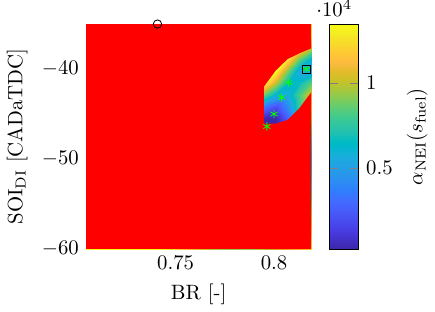}
        \caption{Iteration $k = 5$} \label{fig:result_nei_10_5}
    \end{subfigure}
    \begin{subfigure}[b]{0.45\textwidth}
        \centering
        \includegraphics{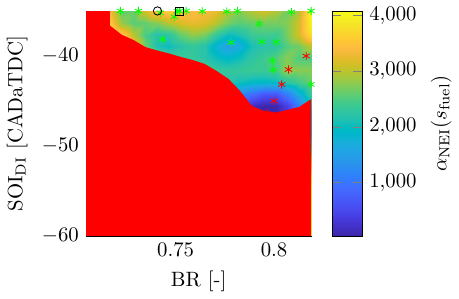}
        \caption{Iteration $k = 25$}\label{fig:result_nei_10_25}
    \end{subfigure}
    
    \begin{subfigure}[b]{0.45\textwidth}
        \centering
        \includegraphics{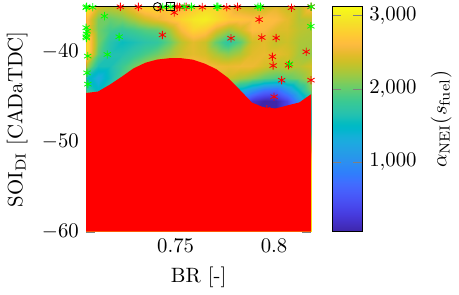}
        \caption{Iteration $k = 50$}\label{fig:result_nei_10_50}
    \end{subfigure}
    \begin{subfigure}[b]{0.45\textwidth}
        \centering
        \includegraphics{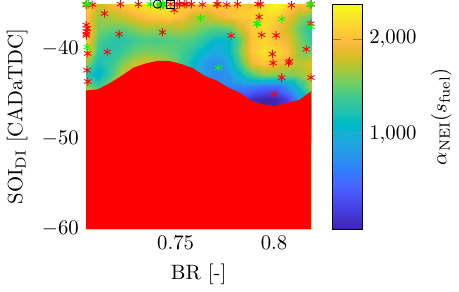}
        \caption{Iteration $k = 68$}\label{fig:result_nei_10_75}
    \end{subfigure}
    \caption{Solving the \acrlong{acr:bo} using $\alpha_\text{NEI}$ with default constraint settings. New samples (green), old sample (red); best observed iteration (square), optimum (circle).} \label{fig:result_nei_10}
\end{figure*}
Figure~\ref{fig:result_nei_10} shows the acquisition function and constraint boundaries at different iterations of the \gls{acr:bo} using $\alpha_\text{NEI}$. It shows in Figures~\ref{fig:result_nei_10_5}~until~\ref{fig:result_nei_10_50} that up to the first 50 iterations the \gls{acr:bo} is expanding the constraint boundaries. The active constraint is the constraint on $\text{IMEP}_\text{g}$ and $\text{cov}(\text{IMEP}_\text{g})$. After 50 iterations the algorithm focusses on refining the information gathered in the area within the constraint bounds as shown in Figures~\ref{fig:result_nei_10_75}.

The other acquisition functions presented in Section~\ref{sec:af} show similar patterns. For brevity, they are discussed in \ref{app:vis_alpha}.

\subsection{Comparison Between Acquisition Functions}
\begin{figure*}
    \centering
    \begin{subfigure}[b]{0.4\textwidth}
        \centering
        \includegraphics{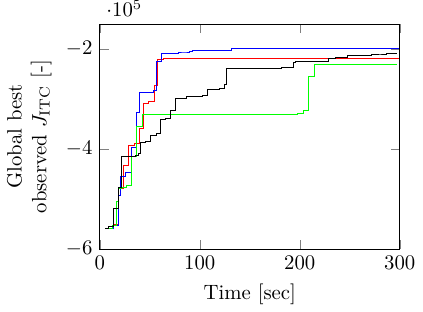} 
        \caption{} \label{fig:compare_time_phi}
    \end{subfigure}
    \hspace{1em}
    \begin{subfigure}[b]{0.4\textwidth}
        \centering
        \includegraphics{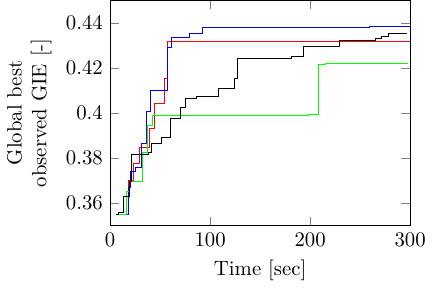} 
        \caption{} \label{fig:compare_time_gie}
    \end{subfigure}
    
    \begin{subfigure}[b]{0.4\textwidth}
        \centering
        \includegraphics{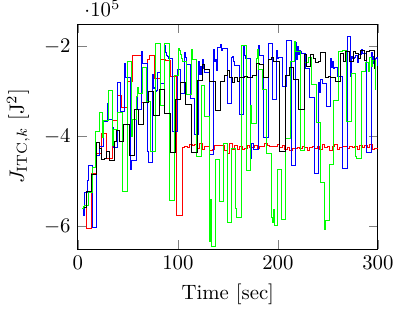} 
        \caption{} \label{fig:compare_time_phi_k}
    \end{subfigure}
    \hspace{1em}
    \begin{subfigure}[b]{0.4\textwidth}
        \centering
        \includegraphics{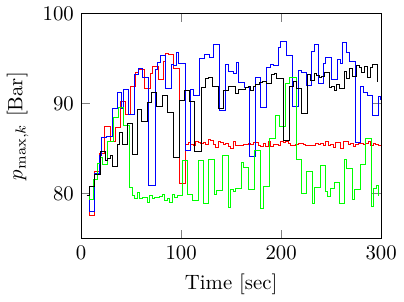} 
        \caption{} \label{fig:compare_time_pmax_k}
    \end{subfigure}
    
    \begin{subfigure}[b]{0.4\textwidth}
        \centering
        \includegraphics{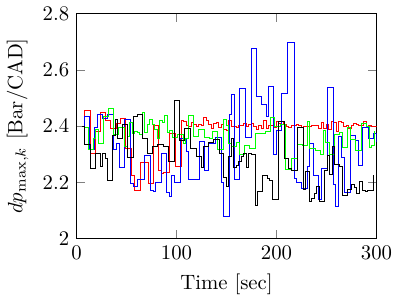} 
        \caption{} \label{fig:compare_time_dpmax_k}
    \end{subfigure}
    \hspace{1em}
    \begin{subfigure}[b]{0.4\textwidth}
        \centering
        \includegraphics{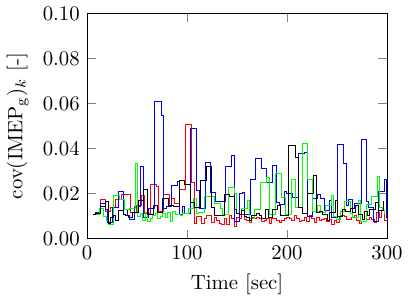} 
        \caption{} \label{fig:compare_time_covIMEP_k}
    \end{subfigure}
    \caption{Important observed parameters over time using $\alpha_\text{EI}$ (red), $\alpha_\text{NEI}$ (blue), $\alpha_\text{PI}$ (green) and $\alpha_\text{NPI}$ (black)} \label{fig:compare_time}
\end{figure*}
\begin{figure*}
    \centering
    \begin{subfigure}[b]{0.4\textwidth}
        \centering
        \includegraphics{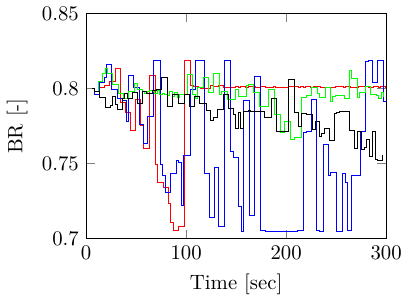} 
        \caption{} \label{fig:compare_time_br}
    \end{subfigure}
    \hspace{1em}
    \begin{subfigure}[b]{0.4\textwidth}
        \centering
        \includegraphics{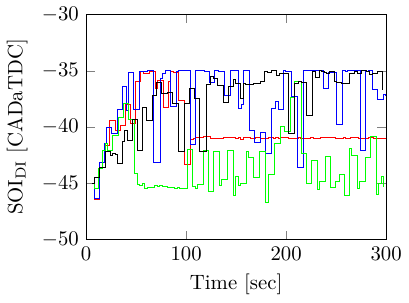} 
        \caption{} \label{fig:compare_time_soi}
    \end{subfigure}
    \caption{Actuator settings over time using $\alpha_\text{EI}$ (red), $\alpha_\text{NEI}$ (blue), $\alpha_\text{PI}$ (green) and $\alpha_\text{NPI}$ (black)} \label{fig:compare_actuator_time}
\end{figure*}

Figures~\ref{fig:compare_time}~and~\ref{fig:compare_actuator_time} show important observed parameters and the actuator settings over time for the different acquisition functions presented in Section~\ref{sec:af}. Figures~\ref{fig:compare_time_phi}~and~\ref{fig:compare_time_gie} show the best observed cost $J_\text{ITC}$ and \gls{acr:gie} over time. Figure~\ref{fig:compare_time_phi_k} shows the cost at each iteration of the \gls{acr:bo}.  It can be seen that by using $\alpha_\text{NEI}$ and $\alpha_\text{NPI}$ the \gls{acr:bo} converges towards similar \gls{acr:gie}; however, using $\alpha_\text{NPI}$ takes substantially longer. The acquisition functions $\alpha_\text{EI}$ and $\alpha_\text{PI}$ start selecting suboptimal solutions after convergence.

Figure~\ref{fig:compare_actuator_time} shows the selected BR and $\text{SOI}_\text{DI}$ over time for the demonstrated acquisition functions. Figure~\ref{fig:compare_time_br} shows that $\alpha_\text{PI}$ and $\alpha_\text{NPI}$ only explore a small region of the allowed range of BR compared to $\alpha_\text{EI}$ and $\alpha_\text{NEI}$.  Figure~\ref{fig:compare_time_soi} shows that all demonstrated acquisition functions explore a similar range of $\text{SOI}_\text{DI}$. The acquisition function $\alpha_\text{NEI}$ keeps exploring the full operating range within constrained bounds after the optimal solution is found. The acquisition function $\alpha_\text{NPI}$ tends to prefer selecting the actuator settings close to the current best found solution, hence being more exploiting. After finding the optimum solution $\alpha_\text{EI}$ and $\alpha_\text{PI}$ tend to exploit a non-optimal actuator setting.

Figure~\ref{fig:compare_time_pmax_k},~\ref{fig:compare_time_dpmax_k}~and~\ref{fig:compare_time_covIMEP_k} show the values related to the constraints of each iteration. It can be seen that in all cases, the \gls{acr:bo} respects the constraints. Since the constraints are not part of the acquisition function, no difference between constraint handling is observed.

Table~\ref{tab:results} shows a comparison between the four acquisition functions. A comparison is made between the location of the found best solution and the true optimum. Also a comparison is made between the difference in $J_\text{ITC}$ and \gls{acr:gie} at the best found solution and the true optimum. Lastly, the convergence time of each acquisition function is given for the used initial conditions. The convergence time is define as the moment that a new best solution is found, but the improvement in \gls{acr:gie} is less than \SI{0.1}{\%} compared to the best found \gls{acr:gie} at $t = \SI{300}{s}$.

The acquisition functions using the expected improvement $\alpha_\text{EI}$ and $\alpha_\text{NEI}$ found actuator setpoints that have an \gls{acr:gie} less then \SI{0.1}{\%} from the true optimum within \SI{60}{s}. The acquisition function using the probability of improvement found an $\Delta \text{GIE}$ of \SI{1.5}{\%} for $\alpha_\text{PI}$ and \SI{0.4}{\%} for $\alpha_\text{NPI}$; however the convergence time of these acquisition functions is substantially longer compared to $\alpha_\text{EI}$ and $\alpha_\text{NEI}$. The fastest convergence time is achieved with $\alpha_\text{NEI}$ and the slowest convergence time with $\alpha_\text{PI}$.
\begin{table*}
    \centering
    \caption{Overview of the best found solution and performance using the different acquisition function presented in Section~\ref{sec:af}.} \label{tab:results}
    \begin{tabular}{r c c c c c}
        \toprule
        Acquisition & $\Delta \text{BR}$ & $\Delta \text{SOI}_\text{DI}$ & $\Delta J_\text{ITC}$ & $\Delta \text{GIE}$ & Convergence \\
        function & [-] & [CAD] & [J$^2$] & [-] & time [s] \\ \midrule 
        $\alpha_\text{EI}$ & \num{0.020} & \num{0.21} & \num{1.2e+03} & \num{-8.2e-03} & \num{57.7}  \\
        $\alpha_\text{NEI}$ & \num{0.0067} & \num{0.0019} & \num{7.7e2} & \num{-9.3e-4} & \num{56.8}  \\
        $\alpha_\text{PI}$ & \num{0.0267} & \num{0.95} & \num{2.4e+05} & \num{-1.5e-2} & \num{208.7} \\
        $\alpha_\text{NPI}$ & \num{0.0139} & \num{0.29} & \num{1.0e3} & \num{-4.2e-3} & \num{193.2}  \\
        \bottomrule
    \end{tabular}
\end{table*}

The acquisition function $\alpha_\text{NEI}$ is selected as the best option. It finds the true optimum and shows a more exploitive behaviour when the optimum is found and the feasible region is fully explored. Furthermore, a convergence time of \SI{56.8}{s} is found given the used initial conditions.

\section{Conclusions and Future Research}
This paper presents an automated, risk-aware calibration method for \gls{acr:rcci} engine actuator setpoints. The proposed method uses constrained \acrfull{acr:bo} as a probabilistic framework to guide the calibration process. The method does not require prior knowledge about the system and during the calibration process knowledge about the system is collected. Using a previously developed method that predicts the in-cylinder pressure between \gls{acr:ivc} and \gls{acr:evo} the proposed method is able to prevent maximum peak pressures and maximum peak pressure rise-rates that exceed specified bounds.

The proposed method was successfully demonstrate in simulations using a validated \gls{acr:rcci} engine model. The method was able to find the optimal fuelling settings within \SI{60}{s} for the used initial condition and acquisition functions. No constraints were violated during the calibration process. As a result, no unsafe setpoints were applied to the simulated system. 

In the future, experimental demonstration of the proposed self-learning calibration method is planned. This will require a reduction in the computation time of the algorithm. Furthermore, the algorithm can not handle transient behaviour. This becomes especially important when the algorithm is used in on-road applications, since it will be unlikely that the engine will reach steady-state behavior. 

\appendix
\section{Visualisation of the Presented Acquisition Functions} \label{app:vis_alpha}
\begin{figure*}
    \centering
    \begin{subfigure}[b]{0.45\textwidth}
        \centering
        \includegraphics{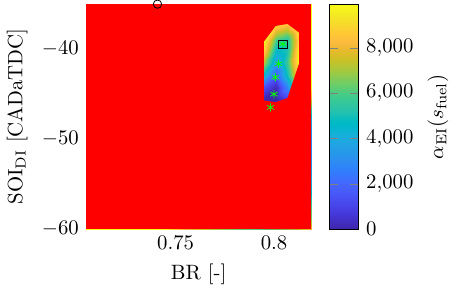}
        \caption{Iteration $k = 5$} \label{fig:result_ei_10_5}
    \end{subfigure}
    \hspace{1em}
    \begin{subfigure}[b]{0.45\textwidth}
        \centering
        \includegraphics{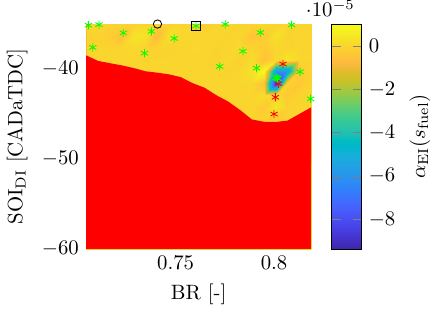}
        \caption{Iteration $k = 25$}\label{fig:result_ei_10_25}
    \end{subfigure}
    
    \begin{subfigure}[b]{0.45\textwidth}
        \centering
        \includegraphics{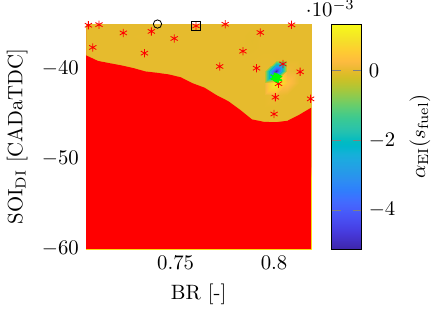}
        \caption{Iteration $k = 50$}\label{fig:result_ei_10_50}
    \end{subfigure}
    \hspace{1em}
    \begin{subfigure}[b]{0.45\textwidth}
        \centering
        \includegraphics{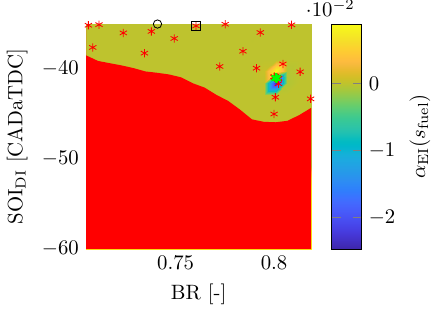}
        \caption{Iteration $k = 75$}\label{fig:result_ei_10_75}
    \end{subfigure}
    
     \begin{subfigure}[b]{0.45\textwidth}
        \centering
        \includegraphics{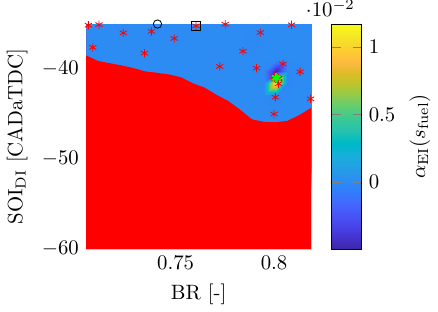}
        \caption{Iteration $k = 100$}\label{fig:result_ei_10_100}
    \end{subfigure}
    \caption{Solving the \acrlong{acr:bo} using $\alpha_\text{EI}$ with default constraint settings. New samples (green), old sample (red); best observed iteration (square), optimum (circle).} \label{fig:result_ei_10}
\end{figure*}
Figure~\ref{fig:result_ei_10} shows the value of the acquisition function and constraint boundaries at different iterations of the \gls{acr:bo} using an acquisition function that does not deal with noisy observations $\alpha_\text{EI}$. Figures~\ref{fig:result_ei_10_5}~and~\ref{fig:result_ei_10_25} show that in the first 25 iterations the optimization with $\alpha_\text{EI}$ explores the constraint boundaries. During this period, the constraint boundary is extended. This increases the feasible operating domain. After the first 25 iterations the optimization with $\alpha_\text{EI}$ selects one point close to the initial guess as shown in Figures~\ref{fig:result_ei_10_50}~and~\ref{fig:result_ei_10_100}. The optimization with $\alpha_\text{EI}$ does not show a tendency to prefer selecting operating points close to the found optimum. 

\begin{figure*}
    \centering
    \begin{subfigure}[b]{0.45\textwidth}
        \centering
        \includegraphics{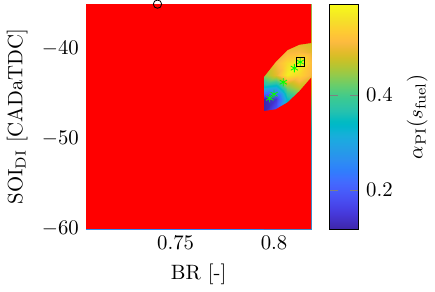}
        \caption{Iteration $k = 5$} \label{fig:result_pi_10_5}
    \end{subfigure}
    \begin{subfigure}[b]{0.45\textwidth}
        \centering
        \includegraphics{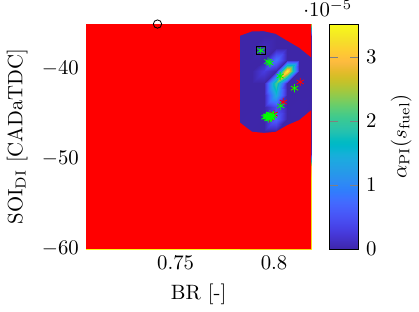}
        \caption{Iteration $k = 25$}\label{fig:result_pi_10_25}
    \end{subfigure}
    
    \begin{subfigure}[b]{0.45\textwidth}
        \centering
        \includegraphics{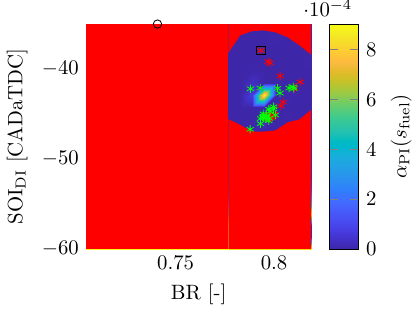}
        \caption{Iteration $k = 50$}\label{fig:result_pi_10_50}
    \end{subfigure}
    \hspace{1em}
    \begin{subfigure}[b]{0.45\textwidth}
        \centering
        \includegraphics{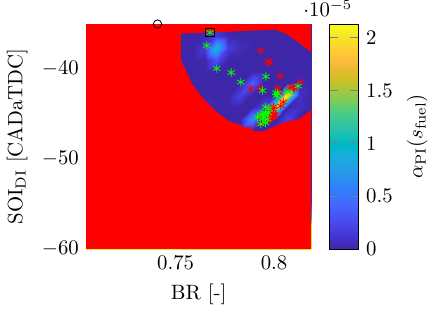}
        \caption{Iteration $k = 75$}\label{fig:result_pi_10_75}
    \end{subfigure}
    \caption{Solving the \acrlong{acr:bo} using $\alpha_\text{PI}$ with default constraint settings. New samples (green), old sample (red); best observed iteration (square), optimum (circle).} \label{fig:result_pi_10}
\end{figure*}
Figure~\ref{fig:result_pi_10} shows the acquisition function and constraint boundaries at different iterations of the \gls{acr:bo} using $\alpha_\text{PI}$. It shows in Figures~\ref{fig:result_pi_10_5}~and~\ref{fig:result_pi_10_25} that up to the first 25 iterations the \gls{acr:bo} is expanding the constraint boundaries. The active constraint is the constraint on $\text{IMEP}_\text{g}$ and $\text{cov}(\text{IMEP}_\text{g})$. After 25 iterations the algorithm focusses on refining the information gathered in the area within the constraint bounds as shown in Figure~\ref{fig:result_pi_10_75}. During these iterations, the constrained boundry is extended a little bit, but is not explored fully. 

\begin{figure*}
    \centering
    \begin{subfigure}[b]{0.45\textwidth}
        \centering
        \includegraphics{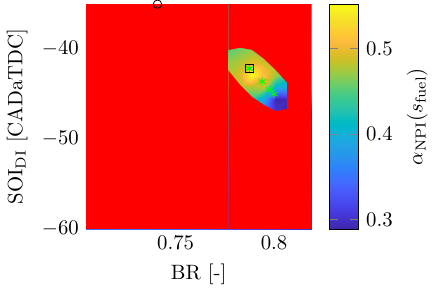}
        \caption{Iteration $k = 5$} \label{fig:result_npi_10_5}
    \end{subfigure}
    \begin{subfigure}[b]{0.45\textwidth}
        \centering
        \includegraphics{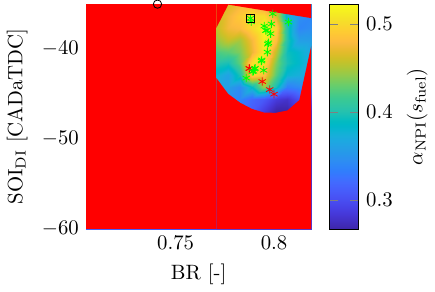}
        \caption{Iteration $k = 25$}\label{fig:result_npi_10_25}
    \end{subfigure}

    \begin{subfigure}[b]{0.45\textwidth}
        \centering
        \includegraphics{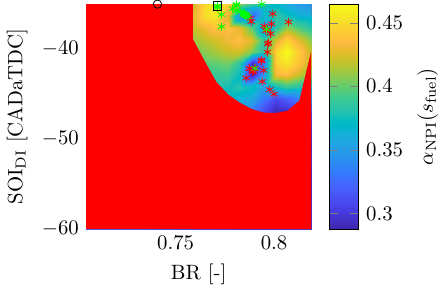}
        \caption{Iteration $k = 50$}\label{fig:result_npi_10_50}
    \end{subfigure}
    \hspace{1em}
    \begin{subfigure}[b]{0.45\textwidth}
        \centering
        \includegraphics{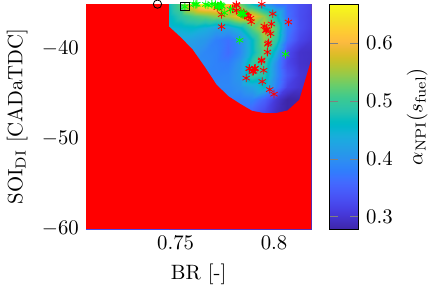}
        \caption{Iteration $k = 75$}\label{fig:result_npi_10_75}
    \end{subfigure}
    \caption{Solving the \acrlong{acr:bo} using $\alpha_\text{NPI}$ with default constraint settings. New samples (green), old sample (red); best observed iteration (square), optimum (circle).} \label{fig:result_npi_10}
\end{figure*}
Figure~\ref{fig:result_npi_10} shows the acquisition function and constraint boundaries at different iterations of the \gls{acr:bo} using $\alpha_\text{NPI}$. It shows in Figures~\ref{fig:result_npi_10_5}~until~\ref{fig:result_npi_10_75} that up to the first 75 iterations the \gls{acr:bo} is expanding the constraint boundaries. This expanding seems to be slower compared to the other acquisition functions. The active constraint is the constraint on $\text{IMEP}_\text{g}$ and $\text{cov}(\text{IMEP}_\text{g})$.

\section*{Declaration of competing interest}
The authors declare that they have no known competing financial interests or personal relationships that could have appeared to influence the work reported in this paper.

\section*{Acknowledgements}
The research presented in this study is financially supported by the Dutch Technology Foundation (STW) under project number 14927.

\bibliography{literature} 

\end{document}